\documentstyle[pre,preprint,aps]{revtex}
\tighten
\input epsf

\begin{document}
\draft

\preprint{SI-97-08}

\title{Universal Short-time Behaviour of \\ 
the Dynamic Fully Frustrated XY Model\thanks{Work supported 
in part by the Deutsche Forschungsgemeinschaft; DFG~Schu 
95/9-1}}

\author{H.J. Luo\thanks{On leave of absence from Sichuan Union 
University,
 Chengdu, P.R. China}, L. Sch\"ulke and B. Zheng}

\address{Universit\"at -- GH Siegen, D -- 57068 Siegen, Germany}

\maketitle

\begin{abstract}
With Monte Carlo methods we investigate the dynamic 
relaxation of the fully frustrated XY model in two dimensions
below or at the Kosterlitz-Thouless phase transition temperature.
Special attention is drawn to the sublattice structure
of the dynamic evolution. Short-time scaling behaviour is found
and universality is confirmed.
The critical exponent $\theta$ is measured for different temperature
and with different algorithms.
\end{abstract}


\section {Introduction}

There has been a long history for the study of universal scaling 
behaviour for critical dynamics. It is well known that there exists
a universal scaling form when dynamic systems almost reach
equilibrium or in the long-time regime of the dynamic evolution.
When a dynamic magnetic system has evolved for a sufficiently long 
time,
the magnetization decays exponentially. The characteristic 
time scale for this regime is 
$t_{\tau} \sim \tau ^ {-\nu z}$ or $t_{L} \sim L^z$ with $\tau$ being
the reduced temperature and $L$ being the lattice size.
Before this exponential decay the time evolution of
the magnetization obeys a power law
 $t^{-\beta/\nu z}$. All this universal scaling behaviour can be
characterized by a set of three critical exponents,
two static exponents $\beta$, $\nu$ and one dynamic exponent $z$.

Is there universal behaviour in the {\it short-time regime }
of the dynamic evolution? For long, the answer had been no.
It was believed that the behaviour of the dynamic systems
in the short-time regime depends essentially on the microscopic 
details. However, it has recently been discovered that
universal scaling behaviour emerges already in the 
{\it macroscopic} short-time regime,
starting at a time just greater than a
 microscopic time scale $t_{mic}$ \cite {jan89}. 
Important is that
extra critical exponents should be introduced 
to describe the dependence of the scaling behaviour on the initial
conditions \cite {jan89,hus89},
 or to characterize the scaling behaviour of
special dynamic observables \cite {maj96,sch97,oer97}.
 A typical dynamic process
is that a magnetic system initially at high temperature 
with a small initial magnetization, is suddenly quenched to
the critical temperature (without any external magnetic
field) and then
released to a time evolution with a dynamics of model A. 
A new dynamic exponent $x_0$ has been introduced to describe
the scaling dimension of the initial magnetization.
More surprisingly, at the beginning of the time evolution
the magnetization undergoes a critical initial increase
\cite {jan89,li94,sch95}
\begin{equation}
M(t) \sim m_0 \, t^\theta
\label{e10}
\end{equation}
where $\theta$ is
 related to the scaling dimension $x_0$ of
the initial magnetization  $m_0$ by
$\theta=(x_0-\beta/\nu)/z$.

The physical background for the initial increase of the magnetization
has not been completely clear. 
The critical initial increase has little to do with
the symmetry breaking below the critical temperature.
Actually similar phenomena can be observed even in case that
the magnetization is not an order parameter.
For the 6-state clock
model and the XY model, below the Kosterlitz Thouless phase
transition temperature no real long range order appears.
The normal magnetization is not an order parameter.
However, the power law initial increase is still observed
in numerical simulations \cite {cze96,oka97}.

On the other hand, in recent years much attention has been drawn to 
statistical systems with frustration or quenched randomness.
Critical behaviour of these systems is often quite different
from that of regular systems. Due to the severe
critical slowing down, numerical simulations of these systems
are extremely difficult. Especially, our knowledge on
the dynamic properties of these systems is poor.

In this paper, as a first approach to the short-time dynamics
of statistical systems with frustrations,
we investigate numerically 
 the dynamic fully frustrated XY model below and at the 
Kosterlitz-Thouless phase transition temperature. We concentrate 
our 
attention on the scaling behaviour of the magnetization
and its dependence on the initial value.
Much effort is devoted to the understanding of the special properties
induced by the sublattice structure of the ground states.
An investigation of the universal short-time behaviour of the
 dynamic systems is not only conceptually important, but also
in the practical sense. Numerical simulations for the Ising
and Potts model show that we may obtain 
already from the short-time dynamics 
the static critical exponents as well as
the dynamic exponent $z$, which are normally defined and measured
in equilibrium or in the long-time regime of the dynamic
evolution \cite {sch95,sch96,gra95,cze96,li95,li96}.
  Since our measurements in the short-time dynamics
are always carried out at the beginning of the time evolution
and the average is really a sample average rather than a time
average based on the ergodicity assumption as in the measurements 
in
equilibrium, the dynamic approach might be free of critical
slowing down.

In the next section, the fully frustrated XY model is briefly
introduced. In Sec.~3 and Sec.~4, numerical results are
reported for the dynamic fully frustrated XY model.
Finally we give some conclusions.

\section {The fully frustrated XY model}

\begin{figure}[p]\centering
\epsfysize=12cm
\epsfclipoff
\fboxsep=0pt
\setlength{\unitlength}{1cm}
\begin{picture}(13.6,12)(0,0)
\put(0,0){{\epsffile{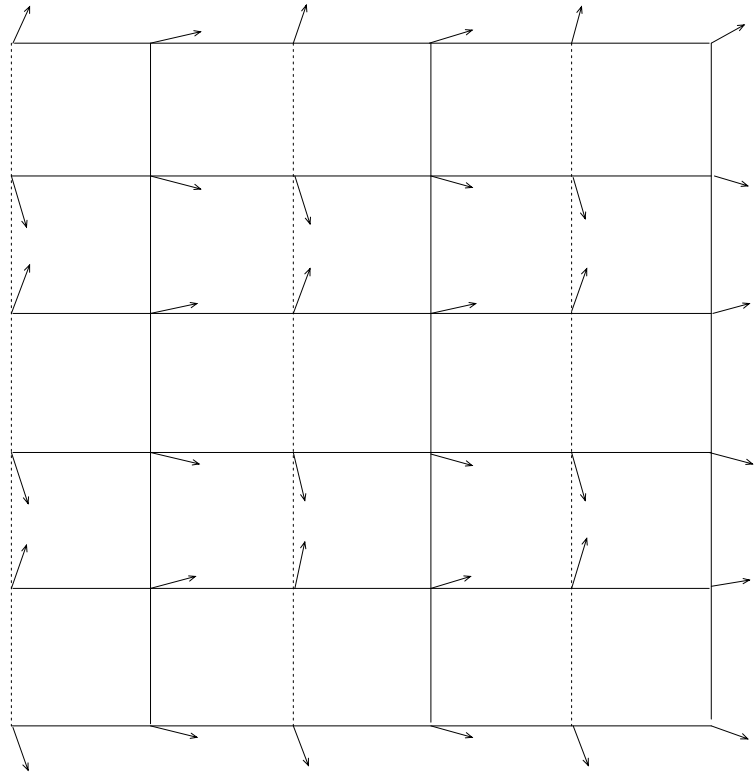}}}
\end{picture}
\caption{ One of the ground states for the FFXY model.
}
\label{f1}
\end{figure}

The fully frustrated XY model in two dimensions (FFXY)
can be defined by the Hamiltonian
\begin{equation}
H=K  \sum_{<ij>} f_{ij}\  \vec S_i \cdot \vec S_j\ ,
\label{e20}
\end{equation}
where $\vec S_i = (S_{i,x},S_{i,y})$ is a planar unit vector at site
$i$ and
the sum is over the nearest neighbours.
 In our notation the inverse temperature
 has been absorbed in the coupling $K$. Here $f_{ij}$ takes
the values $+1$ or $-1$, depending on the links.
A simple realization of the FFXY model is defined by taking
$f_{ij}=-1$ on half of the vertical links (negative links)
while $+1$ on the others (positive links). 
This is shown in Fig.~\ref {f1}.
 The links marked by dotted lines represent 
the negative links.

In the FFXY model two phase transitions exist, 
the Kosterlitz-Thouless phase transition (XY-like) and
the second order phase transition (Ising-like). This is very 
different from the regular XY model. Much effort has been made
to locate the critical points for both transitions and
measure the corresponding
critical exponents. In a recent paper \cite {ols95}, it is reported
that the XY-like phase transition temperature is
$T_{KT}=1/K_{KT}=0.446$ while the Ising-like
 phase transition temperature
$T_{c}=1/K_{c}=0.452$. These two 
measured values for the phase transition temperature
are slightly different. Earlier results for the FFXY model
and its ordering dynamics
can be found in references
 \cite {sch89,lee91,nic91,ram92,lee94,lee95}.

One of the most important properties of the 
fully frustrated XY model is the sublattice structure of its
ground states. In the regular XY model,
in the ground state all spins orient in the same direction.
However, for the FFXY model
 the lattice should be divided into
four sublattices. All spins in each sublattice orient in
the same direction,
but the directions for the four sublattices are different
and connected in a certain  way,  as it is shown in Fig.~\ref {f1}.
For the whole lattice, the ground state  preserves the
global O(2) symmetry at any temperature,
 i.e. spins can rotate {\it globally}.
 No real long range XY-like order 
emerges in the FFXY model. This situation is similar as for the 
regular XY model. Besides the ground state shown in Fig.~\ref {f1},
there is another ground state which is just obtained
by translating the configuration in Fig.~\ref {f1}
by one lattice spacing in
$y$ direction. Below the Ising-like critical temperature $T_c$, 
the $Z_2$ symmetry is broken. A second order 
phase transition occurs.

In this paper we study the short-time 
dynamic properties of the FFXY model
below or at the XY-like phase transition
temperature $T_{KT}$. Below the critical temperature $T_{KT}$,
the FFXY model remains critical in the sense that the correlation
length
keeps divergent. Therefore, a scaling form is expected
even below the critical temperature $T_{KT}$.
However, the critical exponents may vary with respect to the
temperature. Such a phenomenon has been observed
in the 6-state clock model and the regular XY model
\cite {cze96,oka97}.
The dynamic properties related to the 
Ising-like phase transition will not be discussed
in this paper.

\section {Sublattice structure of dynamic evolution}

Let us consider a dynamic relaxation process starting from
an initial state with a very high temperature and small
magnetization \cite {jan89}. As a direct generalization of the XY 
model,
the magnetization for the FFXY model may also be defined as
\begin{equation}
\vec M(t) = \frac {1}{L^2} \sum_{i} \vec S_{i}
\label{e30}
\end{equation}
with $L$ being the lattice size. As in the
XY model, the magnetization here is also not an order parameter.
To achieve an initial magnetization, we introduce 
an initial external magnetic field, e.g. in the $x$ direction, as
in the numerical simulation of the XY model \cite {oka97}.
Taking into account that the initial state is at a very high temperature,
the initial Hamiltonian 
 can be written as
\begin{equation}
H_{0}=2 h  \sum_{i} S_{i,x}.
\label{e40}
\end{equation}
To prepare the initial state, we update the system described
by  the initial Hamiltonian $H_{0}$
until it reaches equilibrium. Then the generated configurations
of this initial system are used as initial configurations
of the dynamic system. The initial magnetization generated is 
\begin{equation}
\vec M(0) = (m_0,0) \approx (h,0), \qquad h \to 0.
\label{e50}
\end{equation}

There are, of course, many other methods to generate 
an initial magnetization.
However, it has been demonstrated that the universal behaviour
does not depend on these microscopic details, i.e. how
a magnetization is constructed \cite {oka97}.
 Effects of the  microscopic details
of the initial configurations are swept away in almost one 
Monte Carlo time step. 

After preparation of an initial configuration,
the system is released to the dynamic evolution
of model A below or at the XY-like transition temperature.
In this paper, the Metropolis algorithm is mainly
used. To confirm universality, however, some simulations
are repeated with the heat-bath algorithm.
We stop updating the dynamic system at $150$ Monte Carlo
time steps, and repeat the procedure with another
initial configuration. The average is over the 
independent initial configurations as well as the random numbers.
We have performed simulations with lattice sizes
$L=8$, $16$, $32$, $64$ and $128$. The total 
number of samples for the average
is between $30\ 000$ and $60\ 000$, depending on lattice sizes,
initial states and algorithms. Errors are estimated
by dividing the samples into three or four groups.

\begin{figure}[p]\centering
\epsfysize=12cm
\epsfclipoff
\fboxsep=0pt
\setlength{\unitlength}{1cm}
\begin{picture}(13.6,12)(0,0)
\put(5.4,8.6){\makebox(0,0){ global start}}
\put(6.,8.){\makebox(0,0){$L=64$, \ $m_0=0.02$}}
\put(0,0){{\epsffile{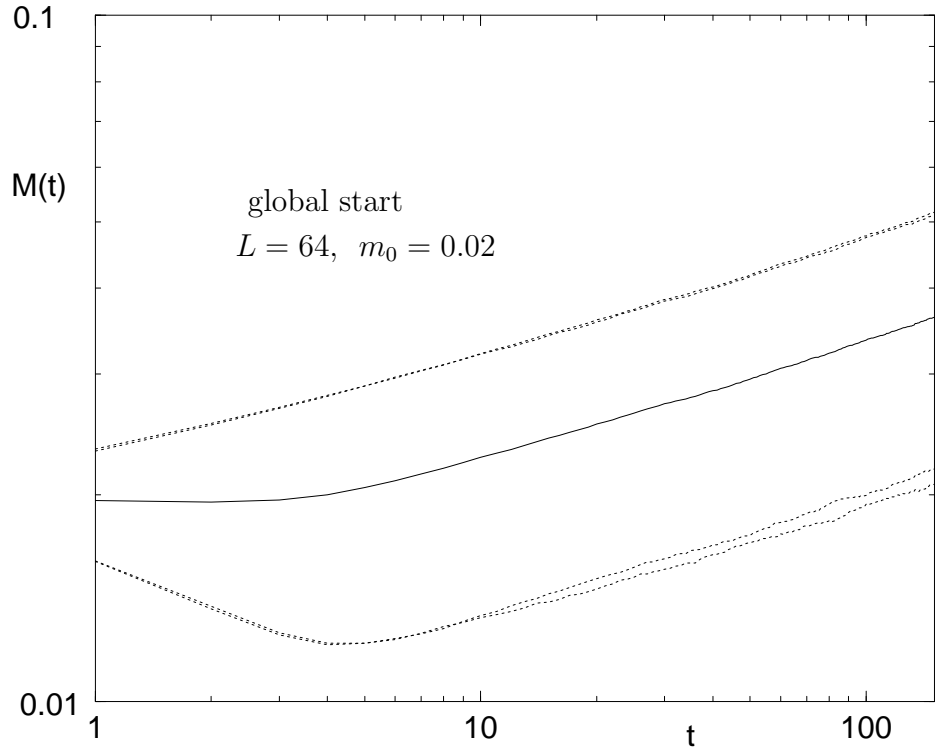}}}
\end{picture}
\caption{ The time evolution of the magnetization with the global start
for $L=64$ and $m_0=0.02$ with the Metropolis
algorithm is plotted in double-log scale.
$M(t)$ is the $x$-component of the magnetization
$\vec M(t)$. The solid line represents the global magnetization
while dotted lines are those for the four sublattices.
}
\label{f2}
\end{figure}

In Fig.~\ref {f2}, the time evolution of the magnetization
is displayed in double-log scale with a solid line
 for the lattice size $L=64$
and the initial magnetization $m_0=0.02$. 
The temperature is taken to be $T=0.400$, which is slightly
below the XY-like transition temperature $T_{KT}=0.446$
given in reference \cite {ols95}
(or $T_{KT}=0.440(2)$ in \cite {lee94}).
In the figure
$M(t)$ is the $x$-component of the magnetization $\vec M(t)$.
The $y$-component of the magnetization $\vec M(t)$ remains zero
since the initial value is zero.
From the figure we see that the magnetization indeed increases
after some time steps. The universal power law behaviour
becomes apparent after about $20 -30$ Monte Carlo time steps.
From the slope of the curve, one measures the exponent
$\theta=0.184(6)$. In this simulation a global uniform
initial external magnetic field $h$ has been applied to the
whole lattice, or in other words, the initial magnetization density
distribution is uniformly generated.
We call this initial state {\it the global start}.

Does the sublattice structure of the ground state play some role
in the dynamic evolution? In equilibrium, it is known that 
the XY susceptibility should be calculated separately for each
sublattice since the orientations of the spins in different
sublattices differ from each other. But further understanding
of the sublattice structure in numerical simulations
can not so easily be achieved due 
to the O(2) symmetry
and large fluctuations. In the short-time dynamics, the situation
is somewhat different. The O(2) symmetry is violated by the 
initial magnetization. One can really measure the time evolution
of the magnetization for each sublattice shown in 
Fig.~\ref {f1} separately.
The results are also included in Fig.~\ref {f2}.
 The two upper dotted lines
represent the time evolution
of the magnetization for the two sublattices connected to
 the positive links while the lower two dotted lines are those
for the two sublattices connected to the negative links.
In is very interesting that for sublattices connected to
 the positive links the magnetization increases from the beginning
while for sublattices connected to
 the negative links they drop significantly in the first time steps.
However, after a certain number of time steps the magnetization 
tends
in all cases 
to the {\it same} universal power law behaviour even though 
the magnitudes remain different for the sublattices
on the positive links and the negative links.

\begin{figure}[p]\centering
\epsfysize=12cm
\epsfclipoff
\fboxsep=0pt
\setlength{\unitlength}{1cm}
\begin{picture}(13.6,12)(0,0)
\put(0,0){{\epsffile{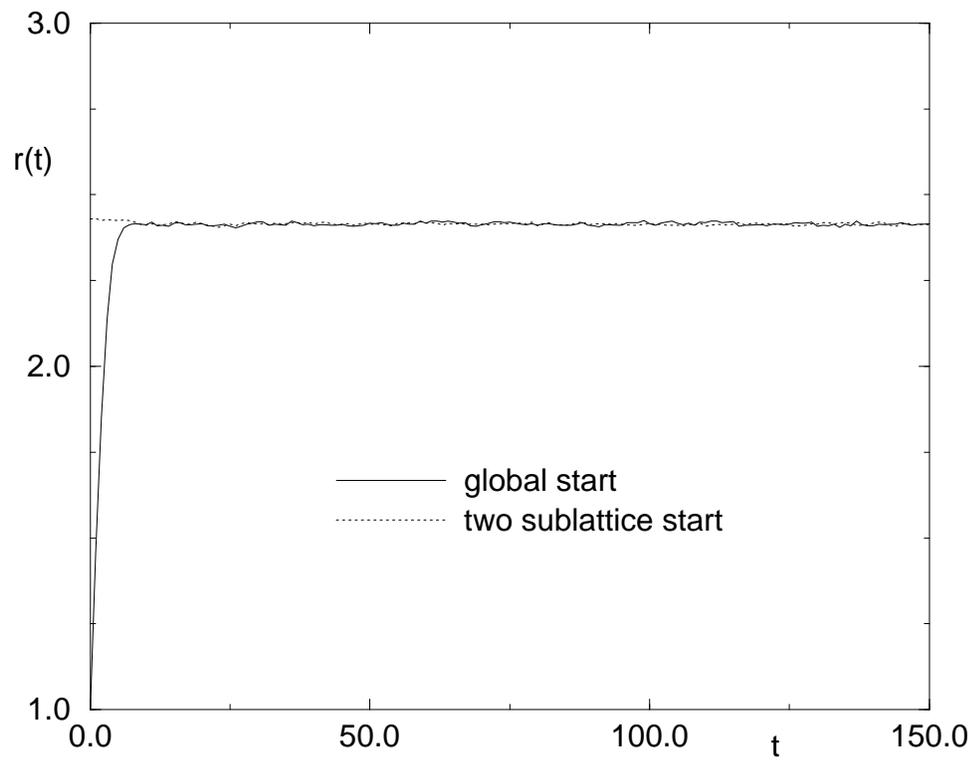}}}
\end{picture}
\caption{ The ratio $r(t)$ for different initial states.
}
\label{f3}
\end{figure}

What is the relation between the magnetizations on the positive 
links and the negative links? In Fig.~\ref {f3}, 
the ratio $r(t)$
of the magnetization averaged over two sublattices on the positive 
links
and that on the negative links is plotted
with a solid line. This ratio 
stabilizes to a constant very quickly within $10$
time steps. Averaging this ratio in a time interval 
$[10,150]$, we get $r=2.4147(8)$.
What is this ratio? In the ground state drawn in Fig.~\ref {f1},
the magnitude of the angles between spins 
on the positive links and the $x$
axis is $\pi/8$, while that between spins on the negative
links and the $x$ axis is $3 \pi/8$. The ratio $r$ should be nothing 
but the ratio
of the $x$-components of the spins on both the positive links
and the negative links. We see this by calculating
$r_{th}=\cos (\pi/8) / \cos (3 \pi /8) \simeq 2.4142$.
The consistence between $r_{th}$ and the measured one 
$r=2.4147(8)$
is remarkable.

\begin{figure}[p]\centering
\epsfysize=12cm
\epsfclipoff
\fboxsep=0pt
\setlength{\unitlength}{1cm}
\begin{picture}(13.6,12)(0,0)
\put(6.1,9.6){\makebox(0,0){ two sublattice start}}
\put(6.,9.){\makebox(0,0){ $L=64$, \ $m_0=0.02$}}
\put(0,0){{\epsffile{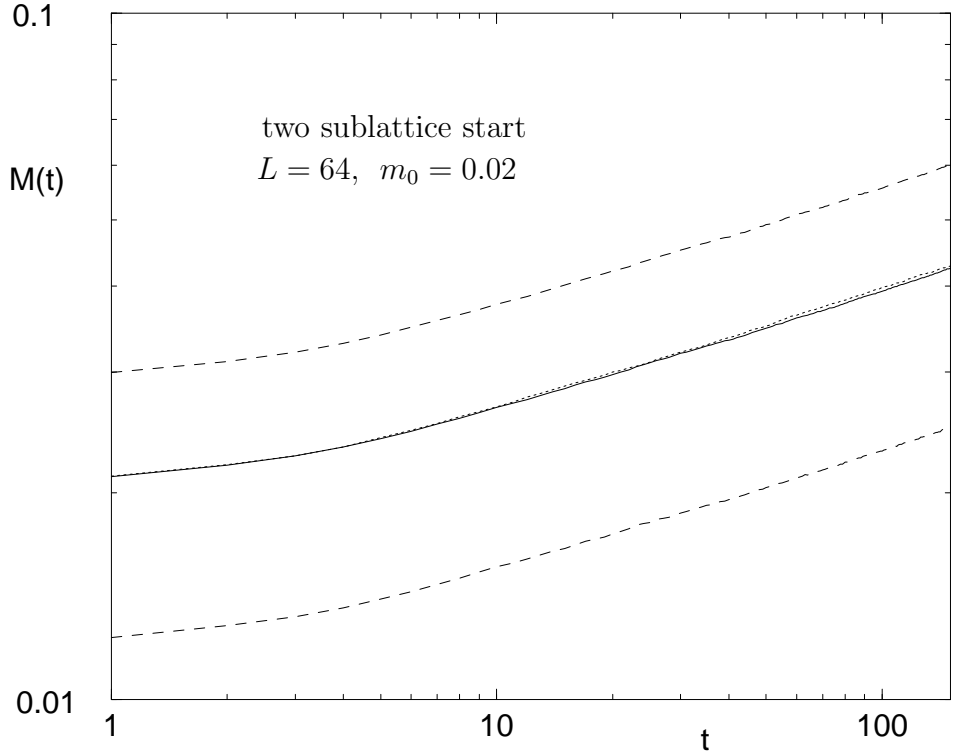}}}
\end{picture}
\caption{ The time evolution of the magnetization with the two 
sublattice start
for $L=64$ and $m_0=0.02$ with the Metropolis
algorithm is plotted in double-log scale.
$M(t)$ is the $x$-component of the magnetization
$\vec M(t)$. The solid line represents the global magnetization
while dashed lines are those for the sublattices. 
For comparison, the dotted line is
the global magnetization with the sharp preparation of
 the initial magnetizations.
}
\label{f4}
\end{figure}

For a better  understanding of this point we now divide
the lattice into {\it two} sublattices, the sublattice on
the positive links and the one on the negative links.
We  have performed
a simulation with different initial magnetizations
for the two sublattices,
$m_{0p}$ and $m_{0n}$ respectively, taking,
for example, $m_{0p}/m_{0n}=r_{th}$
and keeping the global initial magnetization $m_0=0.02$. 
We call such an initial state {\it two sublattice start}.
The time-dependent magnetizations with this initial condition
are plotted in Fig.~\ref {f4}. The solid line is the global
magnetization while the upper and lower dashed line represent
those for sublattices on the positive links and
negative links respectively. Now, the 
decline for the magnetization
on the negative links has disappeared. The corresponding ratio 
$r(t)$ is plotted in Fig.~\ref {f3} with the dotted line.
It is a constant from almost the very beginning of the time
evolution. The averaged value is $r=2.4146(6)$.
 However, the time period for the magnetization
 to enter the universal power
 law behaviour is again $20 - 30$, 
almost the same as  that in Fig.~\ref {f2}.

Before going further we would like to
 mention the problem of the {\it sharp preparation }
of the initial magnetization.
If the lattice size is infinite,
in each initial configuration generated by the initial
Hamiltonian $H_0$ an exact value $(m_0,0)$ of the initial
magnetization $\vec M(0)$ is automatically achieved.
However,
in practice the lattice size is finite and the 
initial magnetization $\vec M(0)$
fluctuates around $(m_0,0)$.
This is a kind of extra finite size effect.
It causes a problem in  high precision
measurements. In order to reduce this effect,
a sharp preparation technique
 has been introduced 
\cite {li94,sch95,oka97a}.
How important the sharp preparation technique is depends
 on the initial magnetization $m_0$ and what kind
of observables one measures. The smaller 
the initial magnetization $m_0$ is, the more important the
sharp preparation technique becomes. In Fig.~\ref {f4},
the result for the time evolution of the
magnetization with the sharp preparation technique
has been plotted with the dotted line. The curve 
almost overlaps with that without the sharp preparation
technique. This shows that the extra finite size effect here
is already quite small for the lattice size $L=64$.
For simplicity, our simulations are always
carried out with no sharp preparation of 
the initial magnetization.

\begin{figure}[p]\centering
\epsfysize=12cm
\epsfclipoff
\fboxsep=0pt
\setlength{\unitlength}{1cm}
\begin{picture}(13.6,12)(0,0)
\put(0,0){{\epsffile{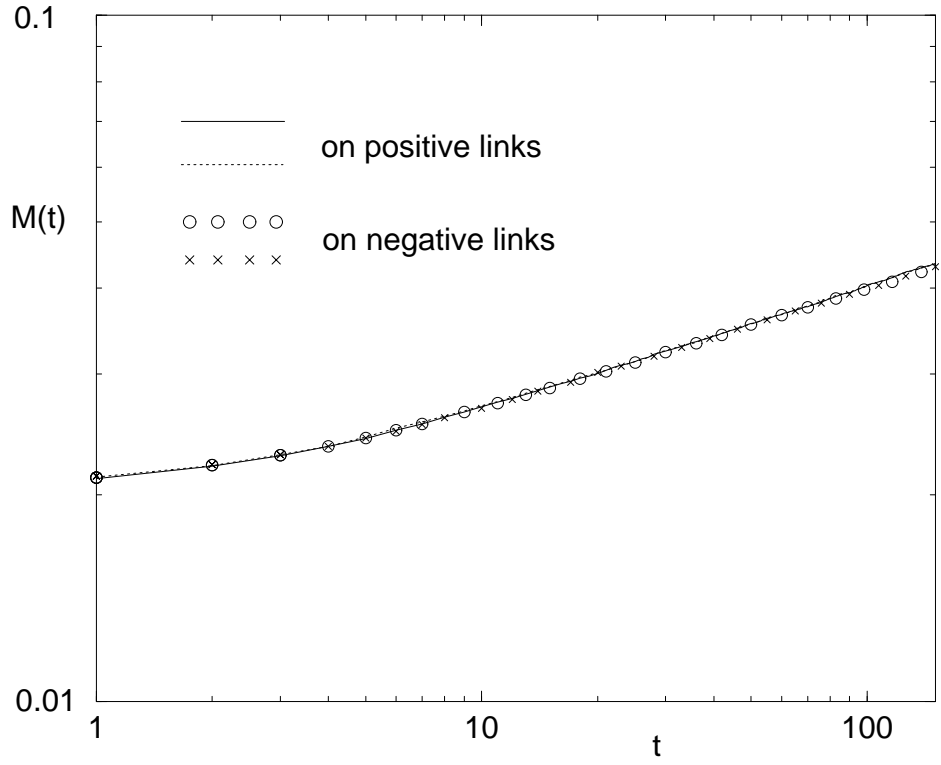}}}
\end{picture}
\caption{ The time evolution of the magnetization with the four 
sublattice start
for $L=64$ and $m_0=0.02$ with the Metropolis
algorithm is plotted in double-log scale.
$M(t)$ is the projection of the magnetization
$\vec M(t)$ on the initial direction. 
The solid and dotted line are the magnetizations
for the sublattices on the positive links
while the circles and crosses represent those on the negative
links.
}
\label{f5}
\end{figure}

Now we go a step further. From the numerical simulations
shown in Fig.~\ref {f2}, \ref {f3} and \ref {f4},
 and the discussions above, 
we understood that the universal behaviour
of the dynamic system has close relation with the
structure of the ground states. If the initial state
is a completely random state with zero initial magnetization,
the probabilities for the magnetization to evolve
to different directions are the same, the averaged
magnetization remains zero. However, if a non-zero initial
magnetization is given to a certain direction,
the time dependent magnetization 
grows in this direction since the energy is
in favour of it. To clarify this point,
we start with an initial state which is even closer to
the ground state. We give different orientations
to the initial magnetizations for the {\it four} sublattices as
 shown in Fig.~\ref {f1}. All the magnitudes of the initial
magnetizations are $m_0=0.02$. This initial state
we call {\it four sublattice start}.
 The time evolution of the
magnetizations for different sublattices are shown in
Fig.~\ref {f5}. Here $M(t)$ is just the projection of the magnetization 
on the direction of the initial magnetization.
The solid line and dotted line are the magnetizations
for the sublattices on the positive links
while the circles and crosses are those on the negative links.
They collapse on a line and show completely the same universal behaviour.

In Table~\ref {t1}, values for the critical exponent
$\theta$ measured in a time interval $[40,150]$
for the three different initial states
are collected. Within the statistical errors they 
are consistent. In the next section
 we will come back to this point.

\section {Universality and scaling}

In the numerical measurements of the critical exponent
$\theta$ we should pay attention to two possible
effects, the finite size effect and the finite $m_0$
effect. 
For the dynamic process discussed in the 
last section, there are two kinds of finite size effects.
One is the extra
 finite size effect from
 the initial configurations. This extra finite size effect
is closely related to the problem
of the sharp preparation of the initial configurations
and has been discussed in the last section.

\begin{figure}[p]\centering
\epsfysize=12cm
\epsfclipoff
\fboxsep=0pt
\setlength{\unitlength}{1cm}
\begin{picture}(13.6,12)(0,0)
\put(5.4,6.9){\makebox(0,0){  solid line:}}
\put(6.,6.4){\makebox(0,0){  $L=64$, \ $m_0=0.02$}}
\put(6.,3.8){\makebox(0,0){  $L=64$, \ $m_0=0.01$}}
\put(0,0){{\epsffile{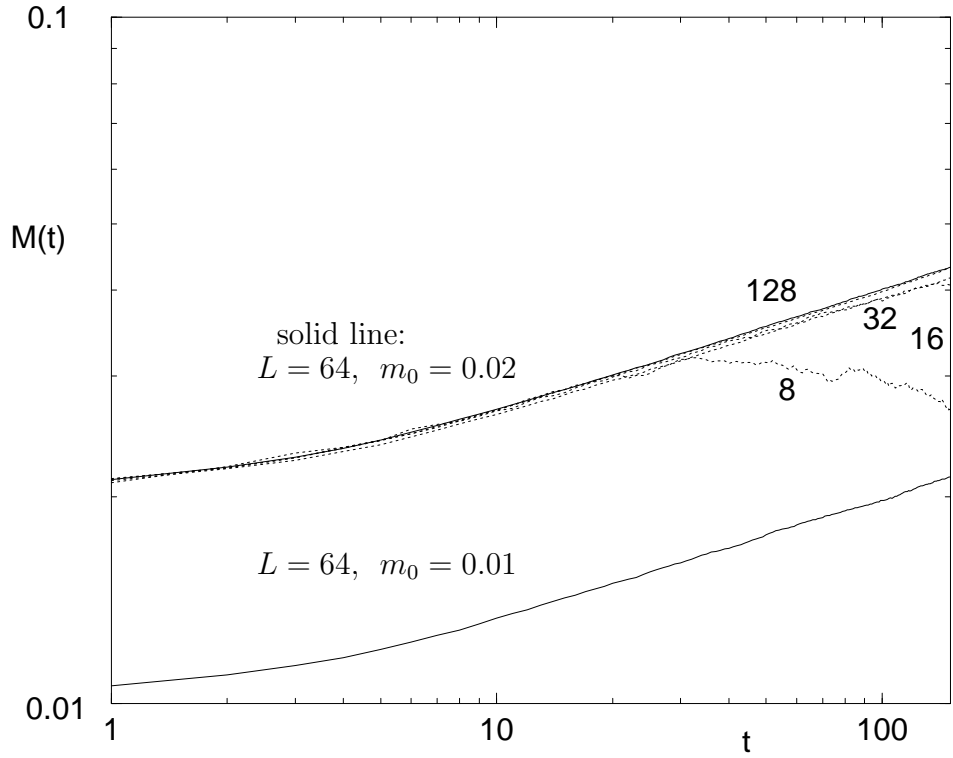}}}
\end{picture}
\caption{ The time evolution of the magnetization with the four 
sublattice start
for different $L$ and $m_0$ with the Metropolis
algorithm is plotted in double-log scale.
$M(t)$ is the projection of the magnetization
$\vec M(t)$ on the initial direction. The solid line represents 
the magnetization
for $L=64$
while the dotted lines are those for other lattice sizes.
}
\label{f6}
\end{figure}

Another kind of finite size effect is the normal 
finite size effect which
takes place in a time scale $ t_L \sim L ^ z$.
Whenever the system evolves into this time regime,
the magnetization will decay by an exponential law
$\exp(-t/t_L)$.
In order to see the normal finite size effect,
we have plotted in double-log scale
the time evolution of the magnetization
for the lattice sizes $L=8$, $16$, $32$, $64$ and $128$
with initial magnetization $m_0=0.02$ in Fig.~\ref {f6}.
The upper solid line is the time-dependent magnetization 
for $L=64$, while the dotted lines are those for 
$L=8$, $16$, $32$ and $128$ respectively. 
The curves for $L=64$ and $L=128$ more or less
overlap. The values for the critical exponent $\theta$
are $\theta=0.181(5)$ and $0.182(3)$ for $L=64$ and $L=128$
respectively.  Therefore we conclude the finite size effect here
is already negligibly small for the lattice size
$L=64$. 

Rigorously speaking, the critical exponent is defined
in the limit $m_0=0$. However, it is practically only possible
 to perform the measurement with finite $m_0$.
The exponent $\theta$ measured 
 may show some dependence
on the initial magnetization $m_0$. In general, a linear extrapolation
to the limit of $m_0=0$ should be carried out
\cite {sch95,oka97a}.
For this purpose, we have performed another simulation
for the lattice size $L=64$ with an initial magnetization 
$m_0=0.01$. The time-dependent magnetization is also displayed
with the lower solid line in
Fig.~\ref {f6}. The exponent $\theta$ obtained for 
$m_0=0.02$ and  $m_0=0.01$ are $\theta=0.181(5)$ and $0.179(7)$
respectively. Within the statistical errors
they cover each other. Since the dependence of the exponent
$\theta$ on $m_0$ is already rather weak here, a
linear extrapolation is not necessary.
This is also the reason why the results with and without
the sharp preparation of the initial magnetization are
not so different. In the further discussions in this paper
we will work mostly with $L=64$ and $m_0=0.02$.

\begin{figure}[p]\centering
\epsfysize=12cm
\epsfclipoff
\fboxsep=0pt
\setlength{\unitlength}{1cm}
\begin{picture}(13.6,12)(0,0)
\put(6.,6.3){\makebox(0,0){  $L=64$, \ $m_0=0.02$}}
\put(8.2,3.7){\makebox(0,0){  $L=64$, \ $m_0=0.01$}}
\put(0,0){{\epsffile{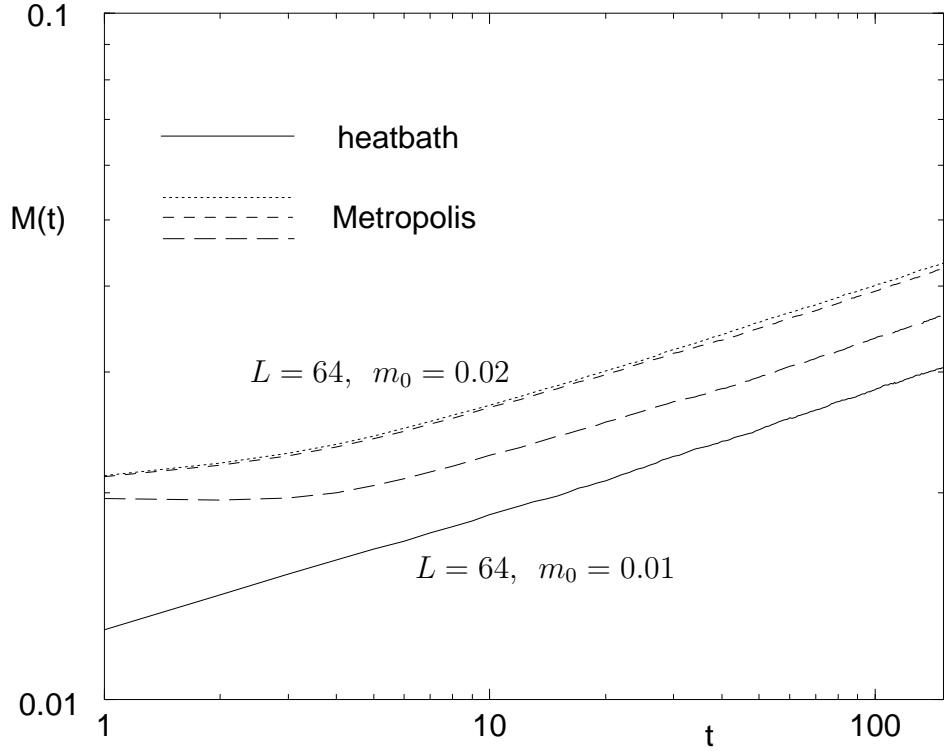}}}
\end{picture}
\caption{ The time evolution of the magnetization
for different initial states and  
algorithms is plotted in double-log scale.
$M(t)$ is the projection of the magnetization
$\vec M(t)$ on the initial direction. The solid line represents the mag-
netization
for $L=64$ and $m_0=0.01$ with the heat-bath algorithm
and the two sublattice start
while long dashed, dashed and dotted lines are those 
for $L=64$ and $m_0=0.02$ with the Metropolis 
algorithm for the global, two sublattice and four sublattice start
respectively.
}
\label{f7}
\end{figure}

Is the power law scaling behaviour in Eq. (\ref {e10})
for the magnetization really universal?
For example, can it depend on the microscopic 
details of the initial state, the algorithms and
the lattice types and even the additional non-nearest
interactions and so on? Many discussions of this kind
have recently been made
 \cite {oka97a,sch97a,oka97,sch97,liu95}. Here we have also
repeated some calculations with the heat-bath algorithm
to confirm universality. The result for
$L=64$ and $m_0=0.01$ with the two sublattice start
is plotted in Fig.~\ref {f7}
with the solid line. The initial magnetization
$m_{0p}$ and $m_{0n}$ are given
to the sublattices on the positive links and the negative
links respectively, with $m_{0p}/m_{0n}=r_{th}=2.4142$.
The long dashed, dashed and dotted line
 are the results with the Metropolis algorithm
 with $m_0=0.02$ and
the global start, two sublattice start and
the four sublattice start respectively.
All the measured values for the corresponding exponent $\theta$
are listed in Table~\ref {t1}.
The fact that all the four curves give consistent
results for the 
exponent $\theta$  within the statistical errors
provides strong support for universality. 
Especially the independence of the exponent $\theta$ on the
initial states indicates that some physical mechanism
closely related to the ground states essentially 
governs the time evolution of the dynamic systems.

To understand universality in the 
short-time dynamics, we keep in mind that there are two
very different time scales in the dynamic systems, 
the  microscopic time scale and the  
 macroscopic time scale. A typical macroscopic time
scale is $t_{\tau}$ or $t_L$. Universal behaviour emerges
only after
a sufficiently long time period in the microscopic sense.
The time period which the dynamic system needs to sweep
away the microscopic behaviour is called $t_{mic}$.
One expects that the time scale $t_{mic}$ is still very 
short in the macroscopic sense. In numerical simulations,
a Monte Carlo time step can be considered as a typical 
microscopic unit. Most of the 
numerical simulations for dynamic systems
show that $t_{mic} \sim 10 - 50$. In Fig.~\ref {f7}
 we see that this is also the case 
for the FFXY model. These results are reasonable.
Compared with the typical macroscopic time scale $t_{\tau}$
ot $t_L$, the microscopic time scale $t_{mic}$ observed
in numerical simulations is indeed very small.
In some cases, universal behaviour emerges actually 
already in one Monte Carlo time step, e.g. in the numerical
measurement of the critical exponent $\theta$ for 
the two-dimensional Potts model \cite {sch95,oka97a}.
Such a clean behaviour in the very short-time regime
is somehow unexpected.

\begin{figure}[p]\centering
\epsfysize=12cm
\epsfclipoff
\fboxsep=0pt
\setlength{\unitlength}{1cm}
\begin{picture}(13.6,12)(0,0)
\put(10.,8.5){\makebox(0,0){  $T=0.250$}}
\put(10.,3.7){\makebox(0,0){  $T=0.446$}}
\put(0,0){{\epsffile{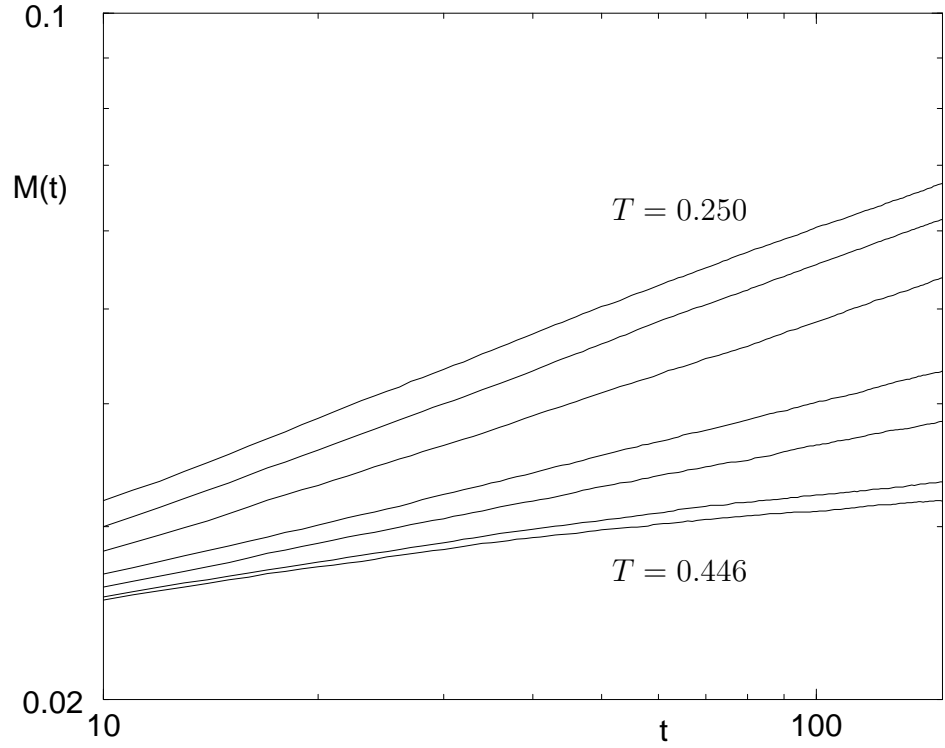}}}
\end{picture}
\caption{ The time evolution of the magnetization
for $L=64$ and $m_0=0.02$ with different temperatures 
is plotted in double-log scale.
The four sublattice start and the Metropolis algorithm are used
in the simulations.
$M(t)$ is the projection of the magnetization
$\vec M(t)$ on the initial direction. From above, the temperature
is $T=0.250$, $0.300$, $0.350$, $0.400$, $0.420$, $0.440$
and $0.446$ respectively.
}
\label{f8}
\end{figure}

It is well known that for the XY-like phase transition
 no real long order emerges even below 
the transition temperature $T_{KT}$. The system remains
critical in the sense that the correlation length is divergent.
A similar scaling behaviour is expected for any temperature below
$T_{KT}$. However, the critical exponents may vary
with respect to the temperature. In Fig.~\ref {f8}, the time evolution
of the magnetization for different temperatures
has been displayed. We clearly see that there exists
power law behaviour for all the temperatures below $T_{KT}$.
However, the critical exponent $\theta$ varies essentially
as can be seen from the results given in Table~\ref {t2}.
The dependence of the critical exponent $\theta$
on the temperature in the FFXY model is qualitatively the same but
stronger than that
for the regular XY model \cite {oka97}.
As the temperature decreases,
the exponent $\theta$ first increases and then slowly decreases.
Around the XY-like transition temperature
$T_{KT}=0.446$ \cite {ols95} (or 0.440(2) in reference
\cite {lee94}), the exponent $\theta$ becomes quite small. Therefore,
the power law behaviour around the critical 
temperature $T_{KT}$ is less prominent compared with that
at the lower temperature. We have also performed some 
simulations with temperatures above $T_{KT}$.
Since the correlation length is divergent {\it exponentially}
when the temperature approaches
the XY-like transition point from above, however, no rigorous 
information for the XY-like transition temperature $T_{KT}$
could be obtained from our data. Further investigation
along this direction is needed.

\section {Conclusions}

We have numerically simulated the dynamic relaxation
process of the fully frustrated XY model (FFXY) in two dimensions
starting from an initial state with a very high temperature
and small initial magnetization. Special attention
has been drawn to the sublattice structure of dynamic evolution
induced by the ground state. Universal power law behaviour
is found independently of the sublattice structure
of the initial states. The critical exponent $\theta$
has been measured for different temperatures
below $T_{KT}$. Universality has been further confirmed 
by carrying out the simulations with both the 
Metropolis and the heat-bath algorithm. Many important
problems remain open in connection with the present
paper, e.g. the short-time behaviour of the dynamic FFXY model
with respect to the Ising-like phase transition or the determination
of the critical point as well as other critical exponents
from the short-time dynamics.

\acknowledgements
One of the authors (H.J. Luo) would like to thank the 
Heinrich Hertz--Stiftung for a fellowship. 


\begin{table}[p]\centering
\begin{tabular}{|c|l|l|l|l|}
        &  \multicolumn{3}{c|} {Metropolis} & {Heatbath}\\
\hline
    & I & II & III & II\\
\hline
 $\theta$ & .184(6) &  .182(5) & .181(5) & .186(6)\\
\end{tabular}
\caption{
 The exponent $\theta$ measured for lattice size $L=64$
with different types of initial configurations and algorithms.
I, II and III represent initial states of global start,
two sublattice start and four sublattice start.
}
\label{t1}
\end{table}

\begin{table}[p]\centering
\begin{tabular}{|c|l|l|l|l|l|l|l|}
 T  & 0.446 & 0.440 & 0.420 & 0.400
    & 0.350 & 0.300 & 0.250\\
\hline
$\theta$ & .060(7) & . 079(4) & .141(5) & .181(5)
       & .245(3)& .263(2) & .260(2)\\
\end{tabular}
\caption{ The exponent $\theta$ measured for different temperatures
with the Metropolis algorithm. The lattice size is $L=64$.}
\label{t2}
\end{table}

\end{document}